\documentclass[twocolumn,showpacs,preprintnumbers]{revtex4}
\usepackage{graphicx}
\usepackage{dcolumn}
\usepackage{bm}

\input{tcilatex}

\begin{document}

\title{Phase diagram of the ferroelectric-relaxor (1-x)PbMg$_{1/3}$Nb$_{2/3}$O$%
_{3}$-xPbTiO$_{3}$}
\author{B. Noheda, D. E. Cox, and G. Shirane}
\affiliation{Department of Physics, Brookhaven National Laboratory, Upton, NY 11973,USA.}
\author{Z.-G. Ye, and J. Gao}
\affiliation{Department of Chemistry, Simon Fraser University, Burnaby, BC, V5A 1S6, Canada.}

\date{\today}

\begin{abstract}
Synchrotron x-ray powder diffraction measurements have been performed on unpoled ceramic 
samples of (1-x)Pb(Mg$_{1/3}$Nb$_{2/3}$)O$_{3}$-xPbTiO$_3$ (PMN-xPT) with 
30\%$\leq $ x$\leq $ 39\% as a 
function of temperature around the morphotropic phase boundary (MPB), which is the 
line separating the rhombohedral 
and tetragonal phases in the phase diagram. The experiments have revealed very 
interesting features 
previously unknown in this or related systems. The sharp and 
well-defined diffraction profiles observed at high and intermediate temperatures in the 
cubic and tetragonal phases, respectively, are in contrast to the broad 
features encountered at low temperatures. These peculiar characteristics, which are 
associated with the monoclinic phase of M$_C$-type previously reported by Kiat et al and Singh et al., can 
only be interpreted as multiple coexisting structures with M$_C$ as the major component. 
An analysis of the diffraction profiles has allowed us to properly characterize the 
PMN-xPT phase diagram and to determine the stability 
region of the monoclinic phase, which extends from x= 31\% to x= 37\% at 20 K. 
The complex 
lansdcape of observed phases points 
to an energy balance between the different PMN-xPT phases which is intrinsically 
much more delicate than that of related systems such as PbZr$_{1-x}$Ti$_x$O$_3$ or (1-x)PbZn$_{1/3}$Nb$_{1/3}$O$_{3}$-xPbTiO$_3$. 
These observations are in good accord with an optical study of x= 33\% by Xu et al., who 
observed monoclinic domains with several different polar directions coexisting with 
rhombohedral domains, in the same single crystal. 
\end{abstract}

\maketitle



\section{Introduction}

Solid solutions of the relaxor-ferroelectrics (1-x)Pb(Mg$_{1/3}$Nb$_{2/3}$)O$%
_3$-x PbTiO$_3$ and (1-x)Pb(Zn$_{1/3}$Nb$_{2/3}$)O$_3$-x PbTiO$_3$, known as PMN-xPT
and PZN-xPT, respectively, are under active consideration for a new generation of
electromechanical devices. When properly oriented, they have piezoelectric 
coefficients which are the highest yet reported, with electromechanical 
deformations one order-of-magnitude larger than those of conventional high 
piezoelectric PbZrO$_{3}$-PbTiO$_{3}$ (PZT) ceramics \cite{Par1,Vie1}. The exceptional 
electromechanical properties of these lead oxide-based solid solutions
have long been known to be related to the nearly vertical phase boundary between the
rhombohedral and the tetragonal phases, the so-called morphotropic phase
boundary (MPB), which is a common feature of the PbMeO$_3$-PbTiO$_3$ systems \cite
{Jaf1,Kuw1,Shr1}.

However, the origin of this unusual behavior remained unknown until
recently, when a monoclinic phase was observed around the MPB of PZT, in
between the rhombohedral and tetragonal phases \cite{Noh1,Noh2}. The monoclinic
phase in PZT, with space group Cm, has the unique axis $b_m$ along the [110]
direction, and a unit cell that is doubled with respect to the primitive
cubic one and rotated $45^o$ about the c-axis with respect to it. We designate
 this monoclinic phase as M$_A$-type, following the notation of Vanderbilt
and Cohen, who in a recent theoretical study derived for the first time a 
region of stability for ferroelectric monoclinic perovskites by an extension of the classic 
Devonshire-Landau expansion of the free energy to eighth-order \cite{Van1}. Diffraction 
experiments with an electric field applied in-situ \cite{Guo1} together with 
first-principles calculations \cite{Bel1} on PZT ceramics have demonstrated a direct 
link between the M$_A$ phase and the high electromechanical deformations in PZT due 
to rotation of the polarization between the [111] and [001] directions.
This fact explains some other studies in which an enhancement of the piezoelectric 
properties of [001]-oriented rhombohedral PZT was reported\cite{Du1,Dam1}

First-principles calculations\cite{Fu1} have also shown that the anomalously
high strain values observed in rhombohedral PZN-xPT crystals under a
[001]-electric field \cite{Par1}can be attributed to polarization rotation from
the rhombohedral [111] to the tetragonal [001] polar axis. Diffraction
experiments with an electric field applied in-situ \cite{Noh4,Ohw1} have
permitted the direct observation of such polarization rotation paths, and have
shown that for rhombohedral PZN-8PT, which has a composition very close to the MPB,
a \textit{different} monoclinic phase (M$_C$ in the notation of ref. \cite{Van1})
is induced when an electric field is applied along the [001] direction. The M$_C$ phase
has the Pm space group, with the unique axis $b_m$ oriented along the pseudo-cubic
[010] direction. With increasing electric field, the M$_C$ phase becomes less distorted and 
approaches tetragonal symmetry but when the field is decreased to zero, an 
orthorhombic (O) phase is reached, similar
to that of BaTiO$_3$, which can be viewed as an M$_C$ phase with $%
a_m=c_m $ \cite{Noh4}. The appearance of this irreversibly-induced O phase
in PZN-8PT suggested that an orthorhombic phase would probably exist
close by in this region of the phase diagram. Shortly afterwards, x-ray powder 
diffraction experiments \cite{Cox1,Lao1} did in fact reveal the existence of such an 
O phase in a narrow composition region (8\%$<$ x $< $11\%) between the
rhombohedral (R) and tetragonal (T) phases in the PZN-xPT phase diagram. The O phase
can be regarded as being very close in energy to the M$_C$ phase for the following two 
reasons; first, a very small elecric field is enough to induce the O-M$_C$ transformation \cite{Noh4} and,
second, one of the four PZN-9\%PT samples studied by Uesu et al. \cite{Ues1}
showed M$_C$ symmetry rather than O.

The MPB phases of the piezoelectric system PMN-xPT\cite{Shr1,Nob1} have not been extensively studied 
until the last two years,  during which a series of papers on this material have appeared. 
PMN-33\%PT single crystals poled along the [011] direction appear to be orthorhombic \cite{Vie2}, while 
X-ray diffraction experiments have shown that a PMN-35\%PT crystal poled along 
the [001] directions has M$_A$-type monoclinic symmetry, similar to that of PZT\cite{Ye1}. This latter  
work also reports that unpoled PMN-35\%PT single crystals are purely
rhombohedral, in contrastr to previous optical microscopy studies of crystals of the 
same nominal composition which reported that they display a complicated pattern of rhombohedral
and tetragonal domains \cite{Ye2}. Very recently, fascinating domain patterns, arising 
from different monoclinic phases with different polarization directions, have been 
observed by Xu et al. in a PMN-33\%PT crystal  \cite{Xu1}. In addition, both 
neutron \cite{Kia1} and x-ray \cite{Sin1} powder diffraction measurements have revealed 
the existence of a monoclinic phase of M$_C$-type, in PMN-35\%PT at low temperatures, and in 
PMN-34\%PT at room temperature, respectively. All these results indicate that the 
behaviour of PMN-xPT around the MPB is very complicated and requires more systematic studies to be properly understood.

In the present work we report a high-resolution synchrotron x-ray powder diffraction 
study on PMN-xPT ceramic samples of several compositions as 
a function of temperature around the MPB. An intermediate monoclinic 
phase of M$_C$ type has been observed in agreement with previous reports \cite{Kia1,Sin1}. 
However, in contrast to these latter results, it has been found that the monoclinic phase does not exist as a single 
phase in any of the samples studied. The present study has allowed us to determine the region of 
stability of the intermediate phase and to construct a new phase diagram for PMN-xPT.

\section{Experimental} 

Samples of (1-x)PbMg$_{1/3}$Nb$_{2/3}$O$_{3}$-xPbTiO$_{3}$ 
with x=0.30, 0.31, 0.33, 0.35, 0.37 and 0.39 (hereafter designated xPT, with x in \%) 
were synthesized 
using the two-step columbite precursor technique \cite{Swa1}. The starting reagents were oxide powders with purities 
better 
than 99.9\%. In the first step, a mixture of MgO and Nb$_{2}$O$_{5}$ containing 
a 15.5 wt\% excess of MgO over 
the 1:1 stoichiometric proportions was thoroughly ground in ethanol and cold-pressed. 
The excess amount of MgO was added in order to compensate for subsequent weight 
losses at high temperatures as determined from a previous DTA/TG analysis. The pellet 
was dried at $80^{o}$ C and calcined at $1100 ^{o}$C for 12h to form pure 
MgNb$_{2}$O$_{6}$ of columbite structure. In the second step, this precursor powder 
was mixed with PbO and TiO$_{2}$ in the appropriate proportions and a 2 wt\% excess of 
PbO added to compensate for the evaporation losses during the calcining and sintering 
processes. Each composition was thoroughly ground and calcined at $900^{o}$C for 4h to 
form the perovskite 
phase. The calcined powders were reground with the addition of a few drops 
of polyvinyl alcohol (PVA) and cold-pressed 
into pellets about 3 mm thick and 15 mm in diameter, which were first heated 
to $650 ^{o}$C in an open Pt crucible for 1h to drive off the PVA, and then 
sintered at $1200 ^{o}$C for 4h in an alumina crucible to form highly densified 
ceramics. The light yellow surfaces of the ceramic pellets were polished with fine 
diamond paste and ultrasonically cleaned. X-ray diffraction patterns obtained on a 
laboratory diffractometer with Cu K$\alpha$ confirmed the formation of perovskite-type 
phases with no evidence of any impurities.

Synchroton x-ray powder diffraction measurements were performed at beamline X7A at the 
Brookhaven National Synchrotron Light Source on several different occasions. In each case a 
double-crystal channel-cut Si (111) monochromator was used in combination with a Ge (220) 
analyzer and a scintillation detector. The wavelength was set to $\approx$ 0.7 \AA\ and 
calibrated with a Si reference standard. With this configuration, the angular resolution 
is better than 0.01$^{o}$. In most cases the data were collected directly from the 
ceramic pellets in Bragg-Brentano geometry by carrying out step-scans at 0.005 or 
0.01$^{o}$ 
intervals over selected angular regions while the sample 
was rocked over a few degrees to improve powder averaging. This type of scan 
provides good 
counting rates but can result in incorrect intensity ratios if significant 
preferred orientation is present in the pellets. However, the peak positions and hence 
the derived unit cell parameters are not affected by this 
problem. Temperature dependence measurements, between 20 and 500K, were performed with 
the 
samples loaded in a closed-cycle cryostat. A further set of measurements was made 
at room temperature on capillary samples which were rotated at about 1 Hz during data 
collection. Such data are free from preferred-orientation 
effects, but suffer from the disadvantage that the counting rates are quite slow. 
For this purpose, a small piece of each pellet was carefully crushed and sieved, and 
the resulting 325-400 mesh fraction (about 38-44 $\mu m$) loaded into a thin 
glass-walled capillary of nominal diameter 0.2 mm \cite{Cox1}. In most cases scans 
were carried out over narrow angular regions centered about the 
six pseudocubic reflections (100), (110), (111), (200), (220) and (222), from which it 
is possible to determine unambiguously the crystal symmetry within the limits of the 
instrumental resolution, together with the corresponding lattice parameters and 
any zero offset if present. For the data analysis, the individual reflection profiles 
were fitted to a pseudo-Voigt function appropriately corrected for asymmetry \cite{Fin1} 
with intensity, peak position, peak-width (FWHM) and mixing parameter as variables. When 
appropriate, the peak widths and mixing parameters were constrained to be the same for 
all the peaks in a clump. Additional peaks were added based on the difference patterns, 
the relative intensities and the goodness-of-fit residuals. Compared to the standard 
Rietveld technique of profile-matching, in which the full diffraction pattern is 
fitted without a structural model, the advantage of this procedure is that it does not 
require any assumptions about the phases to be included, how the FWHM and mixing 
parameters vary as a function of 2$\theta$, and how to allow for anisotropic 
broadening, when neighboring peaks have significantly different widths. We have found that this is a 
fairly common effect in PbO-based perovskite systems, and is likely to occur when one or more of the lattice parameters is more sensitive 
to compositional fluctuations than the others. If overlooked, anisotropic broadening can 
be mistakenly interpreted as a lower-symmetry structure in a profile-matching 
analysis. 

Some information about the microstructure and compositional fluctuations in the samples 
can be obtained from Williamson-Hall plots \cite{Wil1} of the peak widths, which can be 
used to provide estimates of the mean coherence length, \textit{L}, and fluctuations 
in the d-spacings, $\Delta $d/d. The observed values of $\Delta $d/d in the cubic phase are 
consistent with long-range compositional fluctuations in the samples of 
typically $\pm $1\% (FWHM), based on the lattice parameters reported by Noblanc 
et al.\cite{Nob1} for compositions in the range 25-40PT.

\section{Temperature dependence}

The temperature evolution of the lattice parameters for 30PT and
39PT is shown in Fig 1. These two compositions are 
located to the left and right sides, respectively, of the MPB in the PMN-xPT phase 
diagram, and the Curie temperatures, T$_C$, are shown as vertical dotted lines. The 
determination of T$_C$, which is usually made from dielectric measurements, is 
difficult in this case due to the broadness of the dielectric constant peak. 
In this paper the T$_C$ values have been adapted from those of Noblanc et al. \cite{Nob1} 
as the average of the two temperatures, T$_m$ (maximum) and T$_d$ (depoling), 
obtained from dielectric measurements on poled samples, 
which produce better-defined dielectric anomalies. Below their respective T$_C$'s of 
about 400 K and 450 K, 39PT is tetragonal down to 20K, while 30PT is first 
tetragonal becoming rhombohedral at about 350K \cite{Nob1}, and remaining so 
down to 20 K. 

\begin{figure}[tbp]
\includegraphics[width=0.5\textwidth] {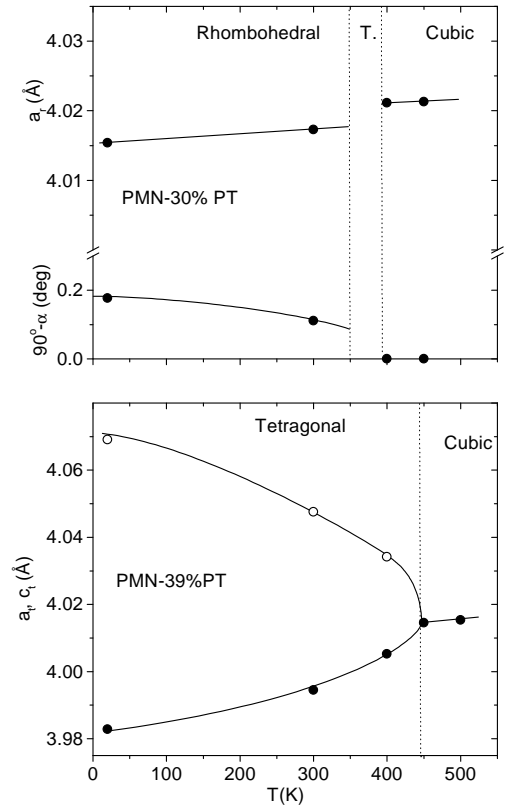}
\caption{Lattice parameters as a function of temperature for the PMN-30PT (top) and PMN-39PT (bottom) 
compositions. Dotted lines represent the respective transition temperatures as defined in the text.}
\end{figure}

The diffraction 
profiles around the pseudo-cubic (111), (200) and (220) diffraction peaks from 
30PT and 39PT at 300K are shown in Fig. 2a and 
Fig. 2d, respectively, in which the least-squares fits to the data points are 
shown as solid lines and the peak positions by vertical arrows. In each case, the 
observed diffraction peaks show no evidence of phase coexistence; 
however, an analysis of the peak broadening reveals that the two compositions 
behave quite differently. Fig. 3 shows the respective Williamson-Hall plots \cite{Wil1}, 
which have slopes of 2$\Delta $d/d and intercepts of $~$$\lambda $$/L$, where $L$ is the coherence length. 
While the 39PT composition shows isotropic broadening with $\Delta $$d/d= 9.5\ \times 10^{-4}$ 
and a coherence length $L=$ 0.5 $\mu $m, the 30PT composition shows highly anisotropic broadening 
and much smaller $L$ values, of about 0.2 $\mu $m, as shown in Fig. 3 (top). In the 30PT composition, the 
broadening is very small for the (hhh) peaks, with 
$\Delta $$d/d= 2.6\ \times 10^{-4}$, and most pronounced for the (h00) peaks, with 
$\Delta $$d/d$ an order-of-magnitude larger. This situation is very similar 
to that observed in the rhombohedral phase of PZT, in which a large degree of intrinsic disorder \cite{Gla1} is known to exist 
due to the existence of local displacements of monoclinic type even though the 
long-range structure is rhombohedral \cite{Cor1}.
 
\begin{figure*}
\includegraphics[width=0.8\textwidth]{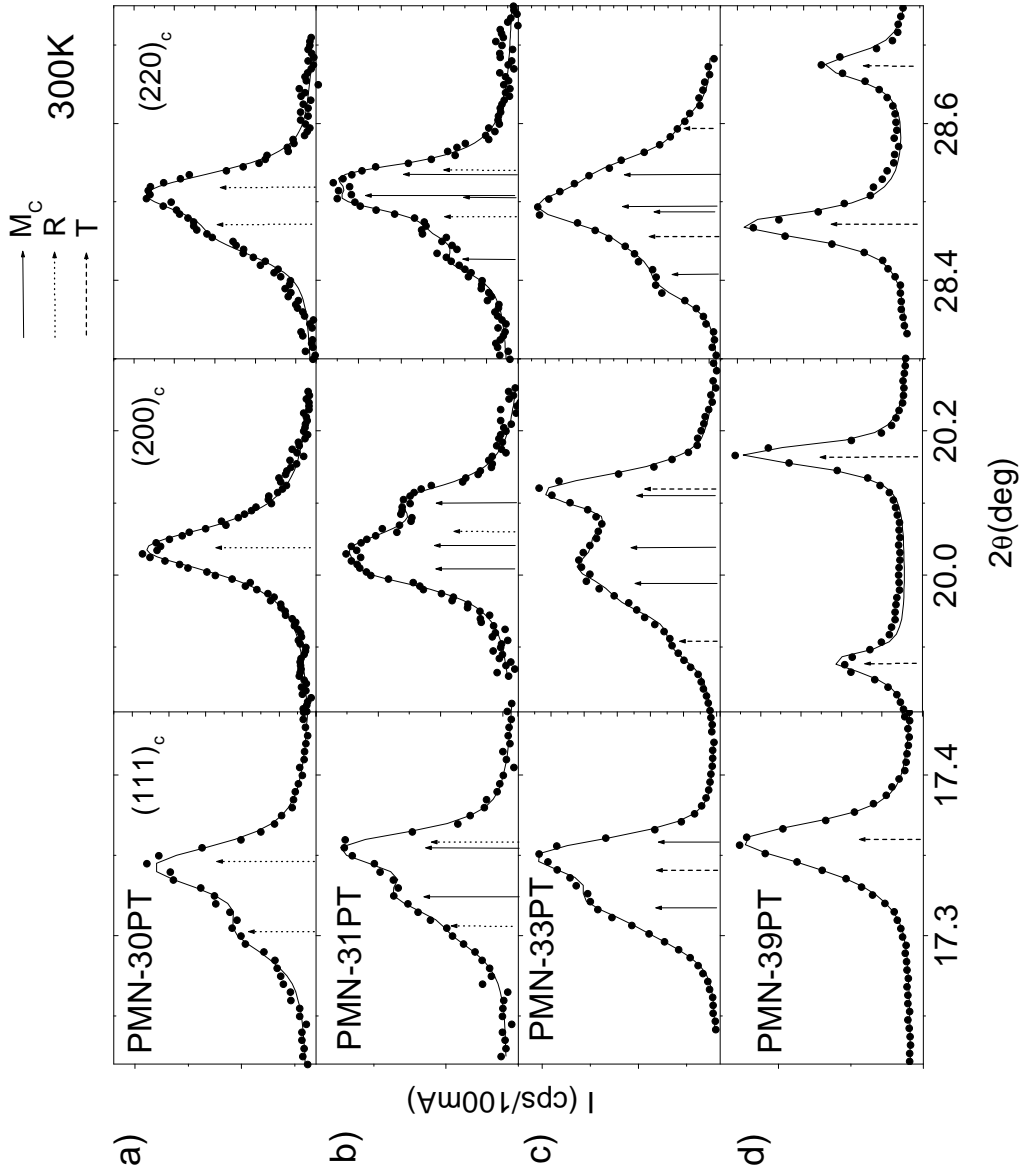}
\caption{Selected regions of the diffraction patterns at 300 K for (a) PMN-30PT, 
(b) PMN-31PT, (c) PMN-33PT and (d) PMN-39PT, corresponding to capillary samples. The solid lines 
are the least-squares fits to the data points and the 
arrows indicate the peak positions obtained from the fits.}
\end{figure*}

Intermediate compositions gave considerably more complicated 
diffraction patterns, as can be seen in 
Figs. 2b and 2c for the 31PT and 33PT samples at 300K. From a detailed analysis of the peak positions and intensities 
of the fitted profiles, it was possible to identify a monoclinic phase of M$_C$-type 
in the 31PT, 33PT, 35PT and 37PT compositions, similar to that reported by 
Kiat et al.\cite{Kia1} and Singh et al \cite{Sin1}. For example, in the case of 
31PT it can be seen in Fig. 2b that the profiles of the pseudo-cubic (200) and (220) 
reflections are distinctly different from those of 30PT shown in Fig. 2a. Various possibilities were considered 
to explain the observed peak positions and intensities of 31PT; 
in particular the lower-symmetry phases previously found in this 
and related systems, i.e. orthorhombic O, monoclinic M$_A$ and monoclinic M$_C$. 
Among these, both O and M$_C$ are found to account quite well for most of the 
observed features, while M$_A$ can be ruled out altogether. 
However, there are several discrepancies in the intensity ratios which can only be 
accounted for by the presence of a minority rhombohedral phase, with an estimated volume 
fraction of about 30\%. 

\begin{figure}[tbp]
\includegraphics[width=0.5\textwidth] {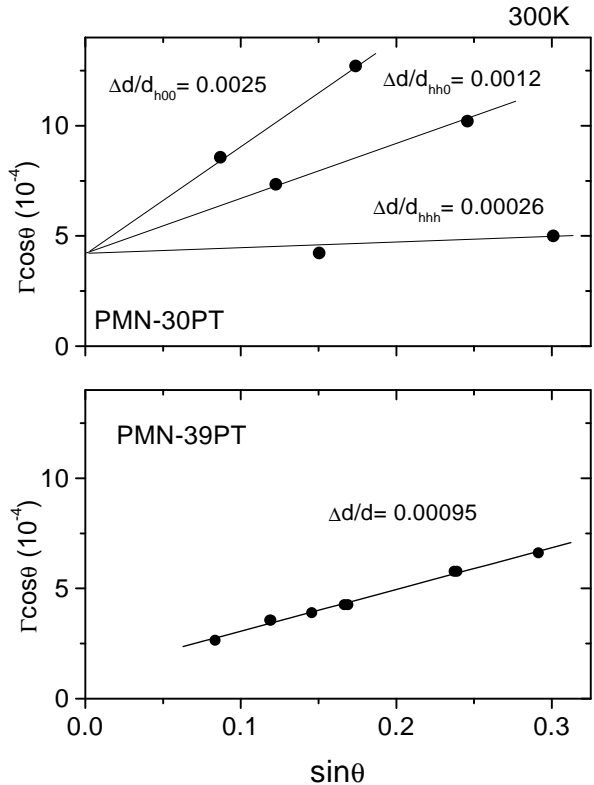}
\caption{Williamson-Hall plots for PMN-30PT (top) and PMN-39PT (bottom) at 300K illustrating the highly anisotropic peak broadening 
in PMN-30PT as compared to 39PT.}
\end{figure}

As mentioned earlier, O and M$_C$ are closely related and very difficult to 
distinguish from each other, since the O phase simply 
represents the limiting case of the M$_C$ phase when $a_m$=$c_m$. In a diffraction 
pattern, the difference between the two is most obvious for the pseudo-cubic 
(200) reflection, which would split into three peaks with roughly equal intensities for 
$M_C$, but only two peaks with a 2:1 intensity ratio for O. In the case of 31PT, it is 
not immediately apparent whether the broad lower-angle peak is composed of one or two 
peaks; however, we were able to obtain a significantly better fit with two peaks, 
which together with the rest of the monoclinic reflections, enabled the lattice 
parameters of the M$_C$ phase to be determined with reasonable accuracy. The 
corresponding peak positions are indicated by solid vertical arrows in Fig. 2b.

\begin{figure}[tbp]
\includegraphics[width=0.5\textwidth] {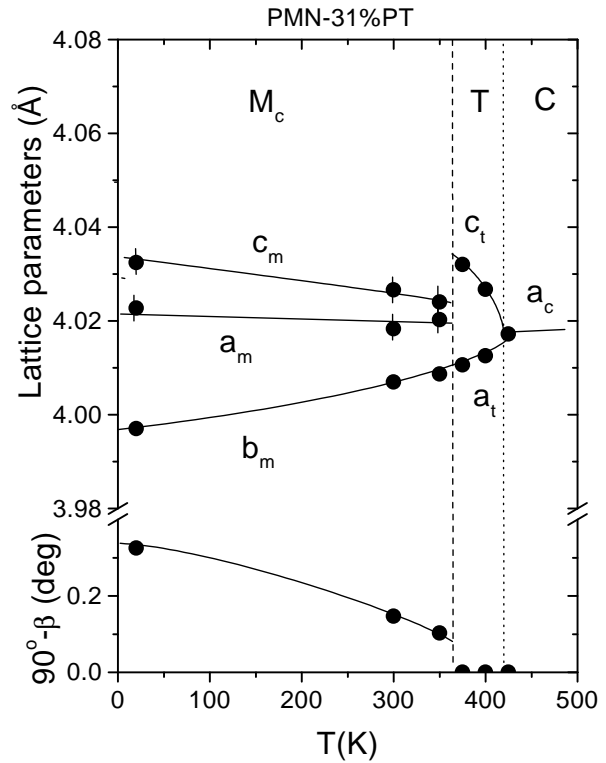}
\caption{Temperature evolution of the lattice parameters of PMN-31PT. The dashed line 
indicates the M$_C$-T transition temperature. 
The dotted line represent the T$_C$ derived from the results of Noblanc et al. 
\cite{Nob1}, as described in the text. }
\end{figure}

The monoclinic lattice parameters of 31PT are plotted as a function of temperature 
in Fig. 4. At 20K, $a_m$ and $c_m$ are fairly well-differentiated, but they approach each other with increasing 
temperature and can no longer be resolved at 350K, as indicated by the error bars in 
Fig. 4. At the same time $b_m$ increases and the monoclinic angle $\beta $, 
approaches $90^o$. Between 350 K and 375 K, the monoclinic phase transforms to a 
tetragonal one, with lattice parameters $a_t$ and 
$c_t$. It is noteworthy that the variation between $b_m$ and $a_t$ is continuous 
at the phase 
transition, in striking similarity to the behavior observed for the orthorhombic 
phase in PZN-xPT \cite{Cox1,Lao1}. 
Finally, between 400K and 425K, the sample transforms into the cubic phase, 
consistent with the PMN-xPT phase diagram in ref. \cite{Nob1}.

\begin{table}
\caption{Lattice parameters at 300K for the PMN-xPT compositions studied. The symmetries 
of the different phases (S) and their volume fractions (f) are indicated in the second and third columns, 
respectively.}
\begin{tabular}{cccccccc}
\hline
x($\%$) & S & f($\%$) & $a($\AA $)$ & $b($\AA ) & $c($\AA $)$ & $\alpha $(=$\gamma $)
$(^{o})$ & $\beta (^{o})$ \\ \hline
30 & R & 100 & 4.017 & 4.017 & 4.017 & 89.89 & 89.89 \\ \hline
31 & R & 30 & 4.017 & 4.017 & 4.017 & 89.89 & 89.89 \\ 
31 & M & 70 & 4.018 & 4.007 & 4.026 & 90 & 90.15 \\ \hline
33 & M & 75 & 4.019 & 4.006 & 4.032 & 90 & 90.19 \\ 
33 & T & 25 & 4.005 & 4.005 & 4.046 & 90 & 90 \\ \hline
35 & M+O& 35 & 4.018 & 4.000 & 4.035 & 90 & 90.12 \\
35 & T & 65 & 4.000 & 4.000 & 4.044 & 90 & 90 \\  \hline
37 & T & 80 & 3.998 & 3.998 & 4.049 & 90 & 90 \\ \hline
39 & T & 100 & 3.994 & 3.994 & 4.047 & 90 & 90 \\ \hline
\end{tabular}

\end{table}

\begin{table}
\caption{Lattice parameters at 20K for the PMN-xPT compositions studied. The symmetries 
of the different phases (S) and their volume fractions (f) are indicated in the second and third columns, 
respectively. The values have been omitted for simplicity.}
\begin{tabular}{cccccccc}
\hline
x($\%$) & S & f($\%$) & $a($\AA $)$ & $b($\AA ) & $c($\AA $)$ & $\alpha $(= $\gamma $) 
$(^{o})$ & $\beta (^{o})$ \\ \hline
30 & R & 100 & 4.015 & 4.015 & 4.015 & 89.82 & 89.82 \\ \hline
31 & R & 30 & 4.020 & 4.020 & 4.020 & 89.85 & 89.85 \\ 
31 & M & 70 & 4.023 & 3.997 & 4.032 & 90 & 90.32 \\ \hline
33 & M & 75 & 4.018 & 3.995 & 4.039 & 90 & 90.32 \\ 
33 & T & 25 & 3.994 & 3.994 & 4.062 & 90 & 90 \\ \hline
35 & M & 40 & 4.011 & 3.990 & 4.049 & 90 & 90.34 \\ 
35 & O & 35 & 4.030 & 3.990 & 4.030 & 90 & 90.34 \\ 
35 & T & 25 & 3.988 & 3.988 & 4.067 & 90 & 90 \\ \hline
37 & M & 55 & 4.015 & 3.985 & 4.039 & 90 & 90.28 \\
37 & T & 45 & 3.989 & 3.989 & 4.062 & 90 & 90 \\ \hline
39 & T & 100 & 3.994 & 3.994 & 4.047 & 90 & 90 \\ \hline
\end{tabular}
\end{table}

At 300K and 20K, the minority rhombohedral phase has the lattice parameters listed in 
Tables I and II. The corresponding peak positions at 300K are 
indicated by the dotted arrows in Fig. 2b. The volume fraction of the rhombohedral 
phase remains constant at roughly 30\% up to 350 K. 

The same regions of the powder diffractogram are plotted in Fig. 2c for the 33PT 
composition at 300K. As in the case of 31PT, identification of the fitted peaks was based 
upon consideration of the R, T, M$_A$, M$_C$ and O phases, both alone and in 
various combinations. Once again, the M$_C$ phase was found to account satisfactorily 
for most of the peaks. Compared to the M$_C$ phase of 31PT (Fig. 2b) there is a larger 
splitting of the three peaks in the pseudo-cubic (200) profile, corresponding to a 
monoclinic distortion with a greater difference between $a_m$ and $c_m$, and hence more 
distinct from the limiting orthorhombic phase. However, as before, there were several 
discrepancies in the intensity ratios and clear evidence of 
a low-angle shoulder on the (h00) profiles which could only be explained by the 
presence of a minority phase, but in this case tetragonal instead of rhombohedral, with 
a volume fraction of about 25\%. The calculated peak positions for the M$_C$ and T 
phases are indicated in Fig. 2c by solid and dashed arrows respectively.

\begin{figure}[tbp]
\includegraphics[width=0.5\textwidth] {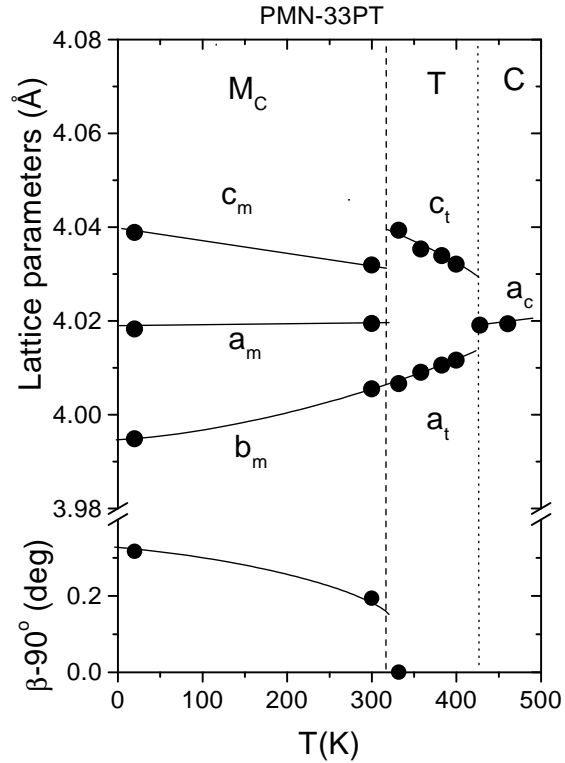}
\caption{Temperature evolution of the lattice parameters of PMN-33PT. The lattice 
parameters of the minority tetragonal phase between 20-300K (see Tables I and II) have been omitted for the sake 
of clarity. The dashed line indicates the M$_C$-T transition temperature. 
The dotted line represent the T$_C$ derived from the results of Noblanc et al. 
\cite{Nob1}, as described in the text.}
\end{figure}

The temperature evolution of the monoclinic lattice parameters is plotted in Fig. 5, and 
is very similar to that of 31PT except for the larger difference between the $a_m$ and 
$c_m$ lattice parameters. Between 300-325K, the monoclinic phase transforms into a 
tetragonal one, which then transforms to the cubic phase 
between 400-425K, in agreement with the previously reported PMN-xPT phase 
diagram\cite{Nob1}. The volume fraction of the minority T phase is also found to be 
constant at about 25\% over the entire temperature range in which it coexists with 
the M$_C$ phase. The $a_t$ lattice parameters at 300K and 20K are not shown in Fig. 5, but 
are very similar to $b_m$ at 300K and 20K, as inferred from the overlap
 of the tetragonal (200) and monoclinic (020) peaks on the high-angle side of the 
(200) pseudo-cubic reflection (see Fig. 2c), and thus follow a continuous trend 
with $a_t$ in the purely tetragonal phase at higher temperatures. $c_t$ follows a similar continuous trend at low temperatures with $c_t$ 
at higher temperatures. The tetragonal lattice parameters for 33PT at 300K and 20K are listed in Tables I and II, 
respectively.

\begin{figure}[tbp]
\includegraphics[width=0.5\textwidth] {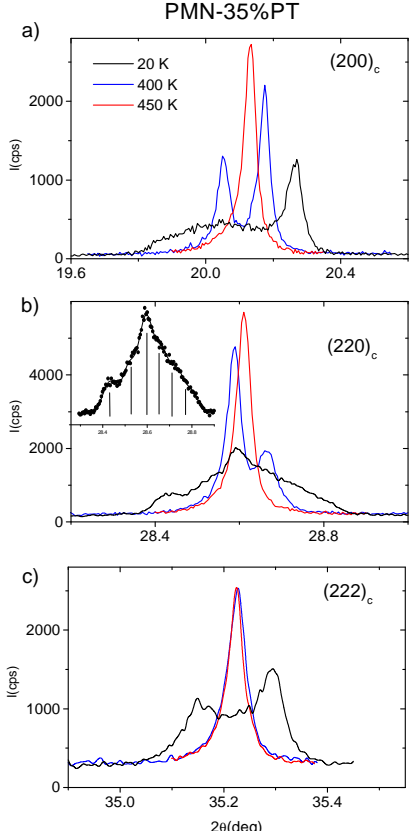}
\caption{(Color) Selected regions of the diffraction pattern of PMN-35PT in the cubic (450K), tetragonal (400K) 
and monoclinic (20K) phases. The inset shows the data points at 20K around (220), 
together with the fit (solid line). The peak positions obtained from the fit are 
indicated by the vertical lines.}
\end{figure}

The behavior observed in 35PT is more complicated. The diffraction profiles around the pseudo-cubic (200), (220) and 
(222) reflections are plotted in Fig. 6 for the three different symmetry regions observed as a function of temperature: the cubic phase 
at 450K, the tetragonal phase at 
400K, and the low temperature region at 20K. The strikingly broad features observed at 
low temperatures are in sharp contrast to the narrow and well-resolved peaks observed 
in the tetragonal and cubic 
phases. The high-angle peak in the (200) profile, which corresponds to the smallest 
lattice 
parameter, $b_m$, shifts towards 
higher angles and remains relatively sharp. In contrast, the intensity in the 
lower-angle region becomes very broad and the fitting procedure 
becomes accordingly more difficult. However, it is quite clear that this 
broadening 
cannot be explained simply by the existence of a second tetragonal phase. Fortunately, 
the 
(hh0) and (hhh) reflections do not show such pronounced broadening and can be fitted 
satisfactorily. The inset to Fig. 6b shows the experimental data points 
around the pseudocubic (220) reflection at 20K together with the solid line representing 
a 
six-peak fit which is in excellent agreement with the observed profile. Most of the 
peak 
positions are consistent with the coexistence of $M_C$ and T phases, and the 
corresponding lattice 
parameters at 300K and 20 K are listed in Tables I and II, respectively. However, 
there remain some significant discrepancies in the intensity ratios and also one 
unexplained peak in the (hh0) profiles. It turns out that these features can be accounted for 
very well by the presence of about 30\% of a third orthorhombic (O) phase with 
$b_m$ and $\beta $ equal to those of M$_C$ phase, and with $a_m$=$c_m$ equal to the mean 
value of the $a_m$ and $c_m$ values for $M_C$ (see table II). At 300K, the O and M$_C$ lattice 
parameters cannot be resolved. 

\begin{figure}[tbp]
\includegraphics[width=0.5\textwidth] {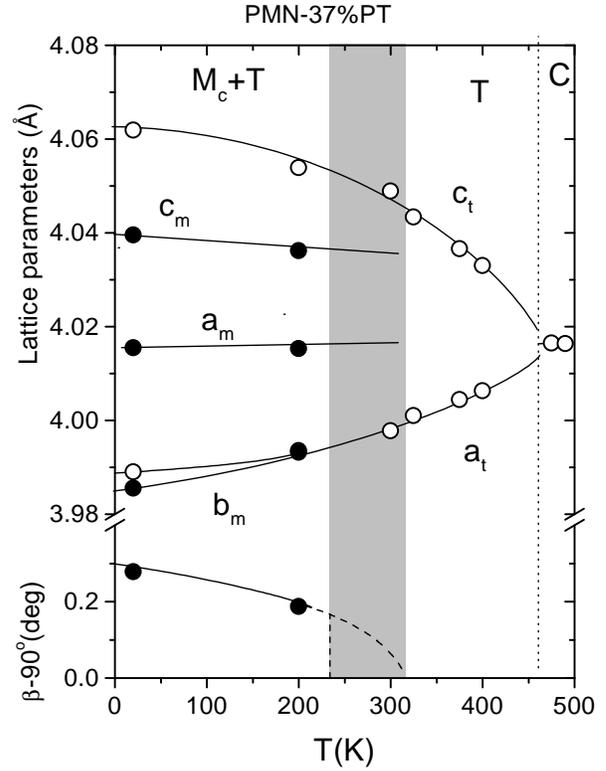}
\caption{Temperature evolution of the monoclinic and tetragonal lattice parameters of PMN-37PT. The 
temperatures interval in which the volume fraction of the monoclinic (tetragonal) phase gradually decreases (increases) is 
shown by the shaded region. The dotted line represent the T$_C$ derived from the results of Noblanc et al. 
\cite{Nob1}, as described in the text.}
\end{figure}

The 37PT composition was found to be a two-phase mixture of M$_C$ and T at low temperatures, 
in the approximate ratio 55:45 at both 20 K and 200K, and predominantly tetragonal at room 
temperature. The evolution of the lattice parameters is plotted in Fig. 7 showing for 
clarity only those phases with a volume fraction in excess of 
40\%. The volume fraction of the T phase increases from about 45\% at 200K, to 80\% at 
300K, and is essentially 100\% at 325K. Thus, as in the case of 35PT, there is a 
temperature interval in which the relative fractions of M$_C$ and T phases are found to 
vary, which has been indicated as a shaded region in Fig. 7. The lattice parameters for 
the T phase at 300K and for the T and M$_C$ phases at 20K, are also shown in Tables I 
and II, respectively. Because of peak overlap, the lattice 
parameters of the minority monoclinic phase at 300K could not be determined. 
The vertical dotted line denotes the temperature of the T-C phase transition obtained, as in 
previously noted, from the mean values of the two temperatures reported in 
ref. \cite{Nob1}.

\section{Composition dependence}

Figs. 8 and 9 show the evolution of the monoclinic phase between the rhombohedral and tetragonal regions 
as a function of composition at 300K and 20K, respectively.  As in previous 
figures, only the values of the majority phases have been plotted for clarity, except 
for 37PT at 20K, where 
the volume fractions of the two phases are roughly the same. The lattice parameters 
reported in ref. \cite{Sin1} for PMN-34\%PT are also plotted in Fig. 8, and are seen to 
be in good agreement 
with the present data. The region of stability of the M$_C$ phase lies approximately 
between 31PT and 35PT at 300K, as 
reported by Singh et al\cite{Sin1}, and it is found to extend slightly further at 20K, to 37PT. The trends in the lattice parameters 
show similar features at both temperatures: $c_{m}$ increases and $b_m$ decreases with increasing Ti content, 
while $a_{m}$ and  ($\beta $-$90^o$) remain approximately 
constant at 300K, and decrease only slightly at 20K. Although there 
are relatively large errors in 
the determination of the lattice parameters, we can nevertheless conclude that the 
transition between the rhombohedral and monoclinic states is rather abrupt, as 
expected on symmetry grounds \cite{Lan1}. More unexpectedly, the transition between the 
monoclinic and tetragonal phases also seems to be fairly abrupt, implying that there is 
no continuous rotation of the polarization all the way to the tetragonal state with 
increasing Ti content. 

\begin{figure}[tbp]
\includegraphics[width=0.5\textwidth] {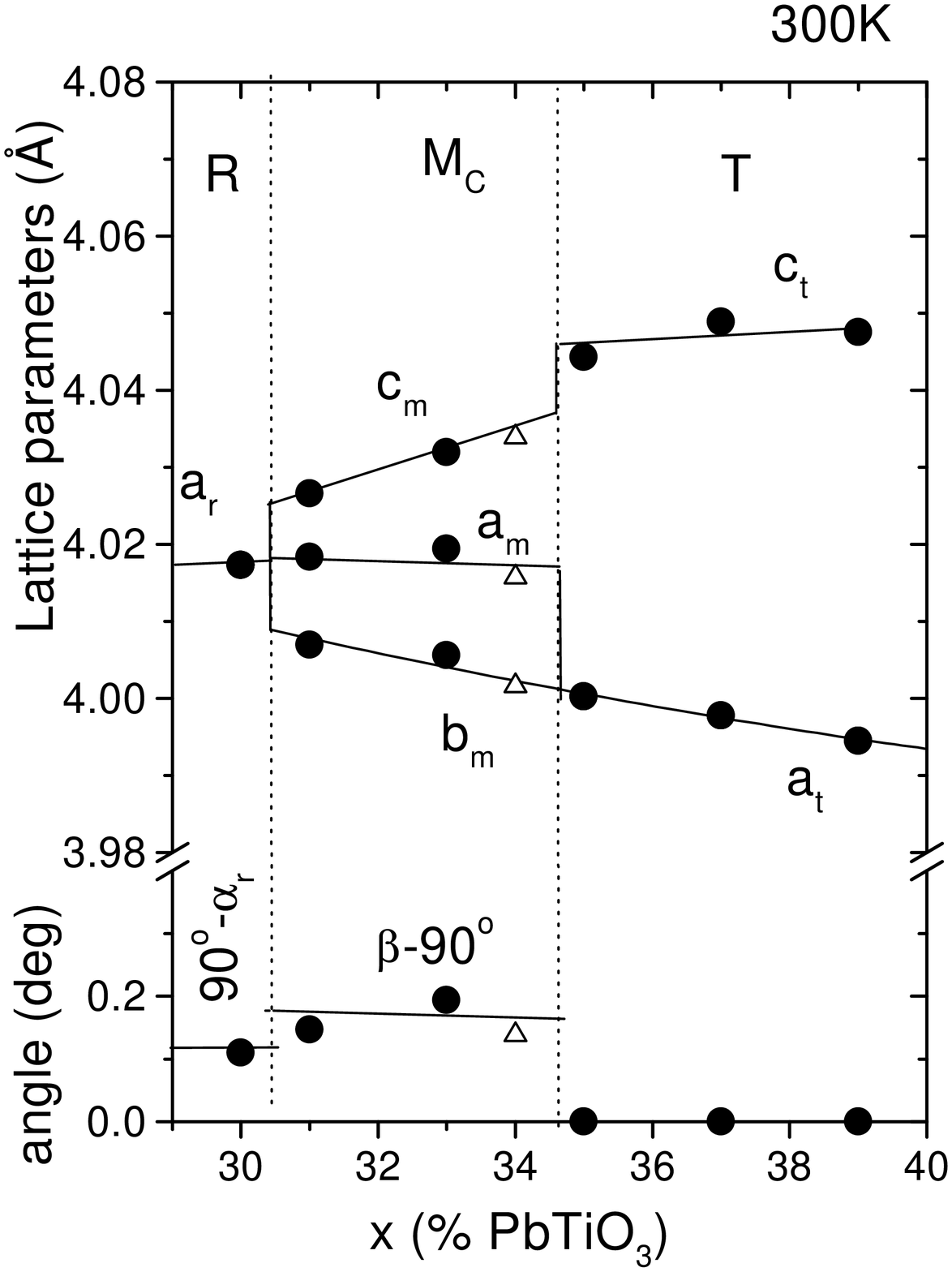}
\caption{Composition dependence of the lattice parameters of the majority phases in PMN-xPT around the MPB at 300K}
\end{figure}

\begin{figure}[tbp]
\includegraphics[width=0.5\textwidth] {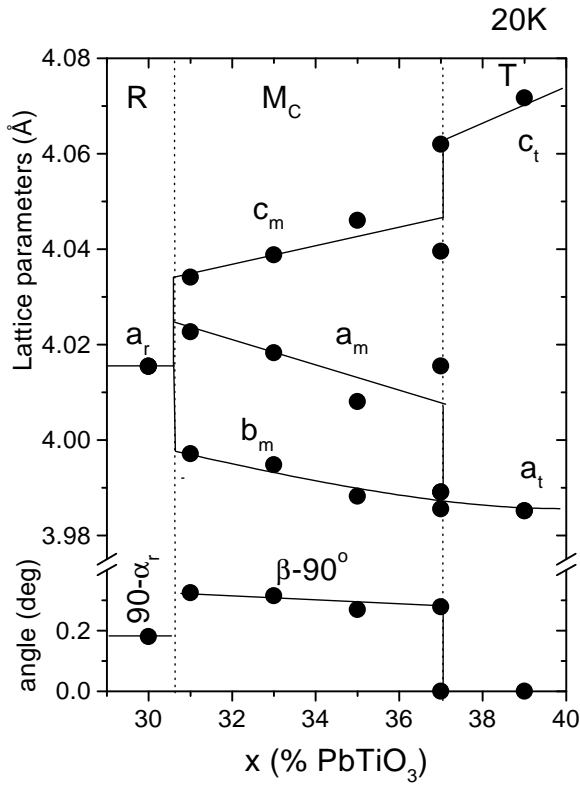}
\caption{Composition dependence of the lattice parameters of the majority phases in PMN-xPT around the MPB at 20K}
\end{figure}

\begin{figure}[tbp]
\includegraphics[width=0.5\textwidth] {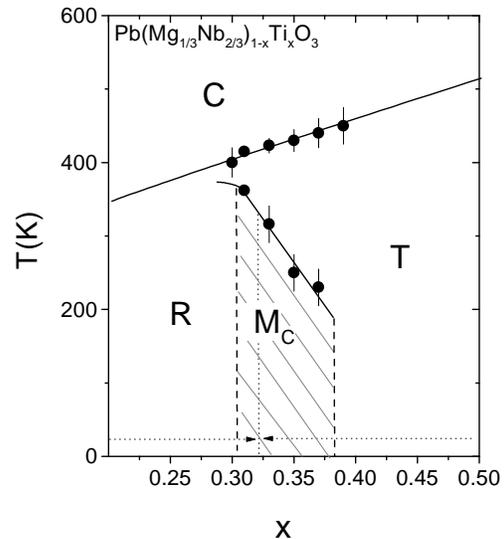}
\caption{Modified phase diagram of PMN-xPT around the MPB. The solid line indicating the 
transition to the cubic phase is the average of the two temperatures reported by Noblanc et 
al.\protect\cite{Nob1} from dielectric measurements. The symbols separating the 
M$_C$ and T phases represent the start of the M$_C$-T phase transition in those cases in which the sample volume 
has been found to transform gradually with temperature (x= 0.35, 0.37)}
\end{figure}

Based on the above analysis, we have constructed a new phase diagram for PMN-xPT in the vicinity of the 
MPB, as shown in 
Fig. 10, in which the stability region of the M$_C$ phase has been shaded. The solid 
line indicating the transition to the cubic phase has been adapted from ref. \cite{Nob1} 
as the average of the two temperatures, T$_m$ (maximum) and T$_d$ (depoling), 
proposed by Noblanc et al. \cite{Nob1}, as previously mentioned. 
These transition temperatures were found to be in better agreement with our data (solid symbols) 
than those reported in ref. \cite{Shr1}. To take into account the observed phase 
coexistence, a dotted vertical line and horizontal dotted arrows have been 
included in the phase diagram, to indicate the rhombohedral ( x$\lesssim$ 32\%) or tetragonal ( x$\gtrsim$ 32\%) symmetry of 
the secondary phase. 

\section{Discussion}

It is interesting to note that the x-ray profiles shown in Fig. 2c for 33PT are clearly similar to those in Fig. 1 of 
ref. \cite{Sin1} for 34PT . In particular, there appears to be definite evidence of a low-angle shoulder in the pseudo-cubic (200) profile, 
and a high-angle shoulder in the (220) profile, both of which would imply the 
presence of a minority tetragonal phase in the latter sample.

Recent work by Topolov \cite{Top1} shows that stress fields may play a very important 
role at the MPB. It  therefore seems plausible that local internal strain due to 
lattice mismatching between coexisting phases is the cause for the observed 
broadening of the pseudo-cubic (h00) profile in the M$_C$ phase. This type of strain 
would produce a distribution of $a_m$ and $c_m$ lattice 
parameters that could explain the striking profiles seen for the 35PT sample at 20K and could 
include the limiting case of an orthorhombic phase for which 
$a_m$=$c_m$. The fact that such a distribution is less evident 
in the other profiles (see Fig. 6b-c) is because the corresponding peak splittings 
are defined mainly by the monoclinic angle and the average values of $a_m$ and $c_m$ 
for (hh0), and exclusively by those values for (hhh). It is also apparent that the 
existence of long-range compositional fluctuations about $\pm 1\% $, indicated by 
Williamson-Hall plots for the cubic phases, as earlier described, can only partially 
explain the observed distribution of monoclinic lattice parameters.

Furthermore, in the context of the three-phase coexistence observed in 35PT at low temperatures, it is 
interesting to note that a degenerate point at which these three phases, T, O and M$_C$ 
coexist has been predicted by Vanderbilt and Cohen \cite{Van1} in an eighth-order 
expansion of the free energy. Although this degeneracy is actually an artifact of the 
model, and could be removed by considering higher-order terms in the Devonshire 
expansion, it is reasonable to suppose that these 
three phases are extremely close in energy and can in fact coexist in real systems due to internal strain, 
thermal fluctuations, etc. The phase diagram reported in ref. \onlinecite{Van1} may 
explain why the MPB of the PMN-xPT system (around the T-M$_C$-O "degenerate" point) is 
considerably more complicated than that of PZT \cite{Noh2} (around the T-M$_A$-R tricritical point) or 
that of PZN-xPT\cite{Lao1} (around the R-O-T tricritical point). Burton et al. \cite{Bur1} have reported a 
slightly lower disorder energy in PMN than in PZN. Moreover, this disorder 
can be expected to increase at the MPB of the PMN-xPT system due to the larger Ti content, 
and could also contribute to the exceptional anharmonicity of the free-energy surfaces \cite{Van1}, a feature believed 
to be related to the nature of the Pb-O bond in these lead oxide solid solutions \cite{Kia2}.

There is the question of whether the observed phase coexistence, R+M$_C$, T+$M_C$, or 
even T+$M_C$+O is an intrinsic feature of the PMN-xPT system. The previously-noted 
degree of compositional fluctuation might by itself explain the coexistence of M$_C$ 
and R phases in 31PT, but not the coexistence of $M_C$ and T over a much wider range. 
According to the eighth-order Devonshire expansion of the free energy \cite{Van1}, a 
phase transition between R and M$_C$ cannot occur directly and is only possible \textit{via} 
an O phase \cite{ref}. In this work, we have observed that 31PT is very close to 
orthorhombic, and it seems entirely plausible that compositions might be found between 
30PT and 31PT which are truly orthorhombic, in which case the transition between 
R and M$_C$ will indeed occur \textit{via} the O phase. 

Furthermore, according to the eighth-order theory\cite{Van1}, the $M_C$-T phase 
transition should be second-order, and no phase coexistence region should be observed. 
However, the O-T phase transition is first-order, and therefore the presence of an O 
phase could explain the wide region of M$_C$+T phase 
coexistence observed. In fact, the existence of a minority 
third O phase, similar to that found in 35PT cannot be neccesarily ruled out in 31PT, 33PT and 37PT, 
although in these cases we are unable to resolve it from the monoclinic phase. 
On the other hand, it is also possible that the continuous character of the M$_C$-T 
transition may be changed by considering higher-orders in the free energy 
expansion \cite{Van1}, since the Landau conditions for second-order transitions that are fulfilled in this case are necessary 
but not sufficient \cite{Lan1}. This would be more in accordance with the fairly abrupt 
changes in the lattice parameters, in special the order parameter $\beta $$-90^o$, 
observed at the transition as a function of composition, and the M$_C$-T phase coexistence could be naturally explained.  

To conclude, a new phase diagram for PMN-xPT has been proposed which delimits the 
the composition range of the M$_C$ phase near the MPB. The existence of a secondary phase, 
either tetragonal or rhombohedral, has been found in all cases (31\%$\leq $ x$\leq $ 37\%). Moreover, the presence of a third minority 
orthorhombic phase, observed for 35PT, cannot be ruled out in the other monoclinic 
compositions, and could explain the observed phase coexistence of M$_C$ and T. This 
complex landscape of phase mixtures, which is in agreement with the observations of Xu et al. for 
33PT single crystals \cite{Xu1}, cannot be explained simply by long-range 
compositional fluctuations in the sample and is believed to be an intrinsic feature of the PMN-xPT system. The evolution of the 
monoclinic lattice parameters as a function of composition shows that the 
M$_C$ phase evolves from the O limit and approaches the T limit, indicating that the 
polarization rotates in the monoclinic (010) plane. However, the rotation appears to be discontinuous at 
both limits. The present results show that the MPB of the MPB system is characterized by multi-phase components and complex phase behavior. 
We hope that our observations will stimulate further experimental and theoretical work 
needed to clarify more precisely the nature of the MPB and its relationship to the electromechanical properties.

\acknowledgments
We would like to thank A. Bokov and M. Dong for their help in sample preparation and  
B. Burton for useful discussions. Financial support from the U.S. Department of Energy under contract 
No. DE-AC02-98CH10886 and  U.S. Office of Naval Research Grant No. N00014-99-1-0738 is also
gratefully acknowledged.


\begin{references}
\bibitem{ref} This would also be possible via a M$_A$ phase. A triclinic phase in between the 
M$_A$ and M$_C$ phase would be likely to exist in this case \protect\cite{Van1,Bel1}.
\end{references}
\end{document}